\documentclass[aps,prd,twocolumn,tightelines,nofootinbib]{revtex4-1}
\usepackage{graphicx}
\usepackage{epsfig}
\usepackage{bm}
\usepackage{latexsym,amssymb,amsmath,amsfonts,amssymb,txfonts,pxfonts,wasysym,float}
\usepackage{mathrsfs}
\usepackage{color}
\usepackage{lettrine}

\usepackage{lipsum}
\usepackage{enumitem}

\usepackage[left,pagewise]{lineno}
\newcommand{\postscript}[2]{\setlength{\epsfxsize}{#2\hsize}
   \centerline{\epsfbox{#1}}}

\usepackage[usenames,dvipsnames]{xcolor}
\definecolor{orange}{cmyk}{0,0.5,1,0}
\definecolor{rossoCP3}{cmyk}{0,.88,.77,.40}
\definecolor{graa}{rgb}{0.8,0.8,0.8}
\definecolor{blaa}{rgb}{0.2,0.2,0.6}

\begin{document}

\title{\color{rossoCP3} El Modelo Est\'andar de la Cosmolog\'{\i}a Moderna en Jaque}

\author{Luis A. Anchordoqui}
\affiliation{Physics Department, Herbert H. Lehman College and Graduate School, The City University of New York\\ 250 Bedford Park Boulevard West, Bronx, New York 10468-1589, USA}

\date{May 2020} 

\begin{abstract}
  \noindent El modelo est\'andar de la cosmolog\'{\i}a moderna permite la
descripci\'on de una amplia gama de datos astron\'omicos y
astrof\'{\i}sicos. Sin embargo, a pesar de este \'exito varias
discrepancias han persistido a lo largo del
tiempo. La m\'as sorprendente es la discrepancia surgida entre la
observaci\'on y los
valores inferidos de la constante de Hubble al usar el modelo cosmol\'ogico. Dicha constante parametriza la
tasa de expansi\'on del cosmos y, por
lo tanto, proporciona pistas sobre el contenido de energ\'{\i}a en el
universo primitivo. En este art\'{\i}culo examinamos el origen de esta
discrepancia y exploramos posibles soluciones para superar el problema.
\end{abstract}
\maketitle

La cosmolog\'{\i}a es una ciencia cuyo objetivo principal es
comprender el origen y la evoluci\'on de las estructuras que
observamos hoy en nuestro universo.
El campo de la cosmolog\'{\i}a ha hecho desarrollos impresionantes en
la \'ultima d\'ecada. Ahora se acepta ampliamente que la cantidad de
materia visible, compuesta mayoritariamente por bariones (protones y
neutrones) y tres
sabores de neutrinos lev\'ogiros (es decir, un estado de helicidad $\nu_L$
junto con sus antineutrinos dextr\'ogiros $\bar \nu_R$), solo forma una peque\~na parte
($\sim 5\%$) de la estructura actual del universo~\cite{Tanabashi:2018oca}. Cabe a\~nadir
que existen s\'olidas evidencias observacionales  que sugieren que la mayor
parte del universo est\'a constituido por las componentes de un sector
oscuro de naturaleza a\'un desconocida.

Si bien se desconoce la naturaleza del sector oscuro, en varios
aspectos es posible hacer estimaciones fundamentadas haciendo uso de
sus efectos gravitacionales. Observaciones astron\'omicas muy
meticulosas han permitido identificar dos tipos de comportamientos
diferentes que se corresponden con la materia oscura, que
constituye el 26\% de la composici\'on del universo, y la
energ\'{\i}a oscura que completa el 69\% restante~\cite{Tanabashi:2018oca}. La evidencia que
respalda la existencia de materia oscura se relaciona principalmente
con anomal\'{\i}as locales en el movimiento de objetos o estructuras
visibles en ciertas regiones del universo. Por otra parte, la
energ\'{\i}a oscura se manifiesta a nivel global con evidencia
observacional que avala una expansi\'on acelerada del universo, la
cual act\'ua de manera opuesta a la gravedad ordinaria de los objetos
masivos, generando un efecto ``antigravitatorio.''

Por medio del efecto gravitacional que el sector oscuro pareciera tener sobre la
materia visible, es posible inferir propiedades sobre el
comportamiento de la materia y la energ\'{\i}a oscura. Es costumbre
suponer 
que la materia oscura es una part\'{\i}cula perfectamente fr\'{\i}a y
estable, y que adem\'as no tiene interacciones. La constante
cosmol\'ogica $\Lambda$ es la suposici\'on m\'as econ\'omica y elegante
que existe en la actualidad para describir el comportamiento de la
energ\'{\i}a oscura~\cite{Peebles:2002gy}. La materia oscura fr\'{\i}a (MOF) y la constante
cosmol\'ogica constituyen la columna vertebral del modelo est\'andar de la
cosmolog\'{\i}a moderna: $\Lambda$MOF. Con estos supuestos podemos estimar como se ver\'{\i}a el modelo
$\Lambda$MOF en el universo primitivo y hacer predicciones para
confrontar con el experimento.

Si bien se sabe que el universo no tiene ning\'un borde, existe
un borde en el universo observable ya que s\'olo podemos ver hasta cierto
punto debido a que la luz viaja a una velocidad finita de alrededor de
300 millones de metros por segundo ($c= 3 \times 10^{8}~{\rm m/s}$). En otras palabras, las se\~nales que nos llegan desde objetos distantes
fueron emitidas hace mucho tiempo atr\'as, por lo que al mirar objetos
distantes estamos mirando hacia el pasado. Suponiendo que el universo
es isotr\'opico, la distancia al borde del universo observable es
aproximadamente la misma en todas las direcciones. Es decir, el
universo observable tiene un volumen esf\'erico (una bola) centrado en
el observador. Cada ubicaci\'on en el universo tiene su propio
universo observable, que puede superponerse o no con el centrado en la
Tierra. El borde del universo observable es tambi\'en llamado horizonte,
ya que proporciona una barrera de lo
que puede ser observado en cada instante de tiempo. M\'as all\'a de
este borde existir\'an part\'{\i}culas cuya luz todav\'{\i}a no ha
tenido tiempo de alcanzarnos, debido a que la edad del universo es
finita. Todo suceso actual o pasado situado por detr\'as de este horizonte de eventos, no forma parte del universo observable actual (aunque puede ser visible en el futuro cuando las se\~nales luminosas procedentes de ellos alcancen nuestra posici\'on futura).

Mirando hacia el espacio profundo y, por lo tanto, viajando
hacia atr\'as en el tiempo, es posible observar que durante los
\'ultimos 13.8 mil millones de a\~nos por donde quiera que miremos existe
una radiaci\'on electromagn\'etica quasi-uniforme e isotr\'opica. Esta radiaci\'on,  que
podemos observar hoy en d\'{\i}a en el rango de las microondas, es
responsable de una peque\~na parte (alrededor del 1\%) de la est\'atica que aparece en las
pantallas de los televisores anal\'ogicos al sintonizar un canal en
blanco. Para comprender el origen de esta radiaci\'on
electromagn\'etica 
es necesario repasar brevemente la historia del universo.

La evidencia observacional indica que el universo comenz\'o con una gran
explosi\'on donde se originaron el tiempo y el espacio, as\'{\i} como
tambi\'en todas las part\'{\i}culas fundamentales (y sus correspondientes antipart\'{\i}culas). Debido a las altas temperaturas, en los primeros momentos despu\'es de la
explosi\'on, el universo era un plasma
caliente, denso y opaco que conten\'{\i}a materia y radiaci\'on
(part\'{\i}culas ``relativistas''  que se mueven a la velocidad de la luz o cerca de ella
como los
fotones y los neutrinos) en interacci\'on. En los primeros tres minutos el universo se expandi\'o lo
suficiente como para que su temperatura bajara desde $10^{32}~{\rm K}$
hasta aproximadamente $10^9~{\rm K}$.\footnote{Para tener en mente una escala
  comparativa, notemos que una temperatura de $10^{9}~{\rm K}
  \equiv$~mil millones de grados por encima del cero absoluto (o grados kelvin), es 70 veces la del interior del Sol.} Este enfriamiento r\'apido
permiti\'o que bajo el efecto de la fuerza nuclear, protones y
neutrones se asociaran para formar n\'ucleos at\'omicos simples a
trav\'es del proceso llamado nucleos\'{\i}ntesis primordial. La fusi\'on de
estos n\'ucleos ocurre aproximadamente antes de los
primeros 20~minutos. Esto se corresponde con el rango de temperatura para que
el plasma primordial fuese lo suficientemente fr\'{\i}o como para que el
deuterio (asociaci\'on de un prot\'on y un neutr\'on) sobreviviera las
colisiones con los fotones, pero a la vez lo
suficientemente caliente y denso como para que ocurrieran reacciones
de fusi\'on a una velocidad significativa que dieron origen a
estructuras cada vez m\'as complejas, como por ejemplo el helio-3 (dos
protones y un neutr\'on) y el helio-4 (dos protones y dos neutrones). La nucleos\'{\i}ntesis primordial ofrece la sonda confiable m\'as
profunda del universo primitivo. Las predicci\'ones de las
abundancias de deuterio y helio-4, est\'an en buen acuerdo con las abundancias
primordiales inferidas de los datos de observaci\'on. 

A medida que el universo se fue expandiendo, el enfriamiento adiab\'atico caus\'o que el plasma se enfriara por debajo de los
$3 \times 10^3~{\rm K}$ permitiendo que protones y electrones se unieran para formar
los primeros \'atomos que constituyen la materia neutra que conocemos
hoy en d\'{i}a (los \'atomos tienen igual n\'umero de carga positiva y
negativa). Esto ocurre alrededor de 380 mil a\~nos ($380$~kyr) despu\'es de la
explosi\'on. A partir de ese momento, al no haber carga el\'ectrica
neta los fotones lograron por primera vez propagarse libremente sin
interactuar con las part\'{\i}culas cargadas que ahora formaban
sistemas neutros (los \'atomos). A medida que el universo se continua
expandiendo, la temperatura de estos fotones disminuye, lo cual explica por que hoy
en d\'{\i}a es de apenas 3~K, correspondi\'endose con el rango de las microondas en el espectro electromagn\'etico.

Desde el momento de la explosi\'on y hasta unos 47~kyr despu\'es, la
densidad de radiaci\'on en el universo era mayor que la densidad de materia. Sin embargo, en un universo en expansi\'on, la densidad de
radiaci\'on disminuye m\'as r\'apido que la densidad de materia,  as\'{i}
que cuando
la temperatura descendi\'o hasta unos 20 mil grados kelvin el universo
comenz\'o a ser dominado por la materia. Claro est\'a que si bien la densidad
de energ\'{\i}a de la materia era mayor que la de la radiaci\'on, la
densidad de energ\'{\i}a en el plasma de bariones y fotones estaba
dominada por la radiaci\'on, por lo que el plasma era relativista. La alta presi\'on de este plasma
provoc\'o oscilaciones. El potencial gravitacional de la materia
oscura era la fuerza impulsora y la presi\'on la fuerza
restauradora. M\'as concretamente, las oscilaciones se produjeron de la
siguiente manera: las regiones de mayor densidad de materia oscura
causaban una mayor atracci\'on gravitacional. En consecuencia, la
densidad del plasma tambi\'en aumentaban en tales regiones. Pero a medida
que el plasma flu\'{\i}a hacia esa regi\'on, se comprim\'{\i}a. El plasma
comprimido ten\'{i}a una presi\'on interna m\'as alta, principalmente
debido a los fotones, pero tambi\'en se transfer\'{\i}a a los bariones a
trav\'es de la interacci\'on electromagn\'etica. Una vez que la
presi\'on hab\'{\i}a aumentado lo suficiente, tend\'{\i}a a alejar a los bariones, lo que conduc\'{\i}a a una densidad de
energ\'{\i}a inferior a la media en esa regi\'on. La presi\'on del
plasma tambi\'en disminu\'{i}a y, por lo tanto, por su atracci\'on
gravitacional, la materia oscura pod\'{\i}a nuevamente atraer m\'as plasma
hacia la regi\'on, lo que aumentaba la densidad del plasma, y ​​el ciclo comenzaba nuevamente. Las oscilaciones de plasma resultantes son
muy similares a las ondas de sonido:  fluctuaciones peri\'odicas de
densidad en el aire. Por lo tanto, a estas oscilaciones del plasma se
las conoce como ``oscilaciones ac\'usticas.'' Es decir, tal como ocurre en el aire,  una peque\~na
perturbaci\'on en la densidad del plasma primordial se habr\'{\i}a
propagado como una onda de presi\'on: un tren de compresiones y
expansiones (o rarefacciones) donde
la presi\'on era m\'as alta o m\'as baja que la media,
respectivamente. Tanto la materia ordinaria como la materia oscura suministraban masa al
gas primordial y por ende ambas generanaban la atracci\'on
gravitacional, pero solo la materia ordinaria sufr\'{\i}a compresiones
s\'onicas y rarefacciones. Las compresiones calentaron el plasma y las rarefacciones
lo enfriaron, por lo que cualquier perturbaci\'on en el universo primitivo result\'o en un patr\'on cambiante de fluctuaciones de temperatura.

Las ondas sonoras viajan a la velocidad del sonido $c_s$. Para ondas
de sonido ordinarias en el aire, esto equivale a alrededor de 300
metros por segundo. Por el contrario, para las ondas de sonido del
plasma en el universo primitivo, la velocidad del sonido equivale
aproximadamente al 60\% de la velocidad de la luz. La velocidad del
sonido nos dice cu\'an r\'apido las perturbaciones de densidad
existentes viajan a trav\'es del espacio. Pero tambi\'en nos dice
cuanto tiempo lleva excitar oscilaciones espec\'{i}ficas: para una
regi\'on de extensi\'on espacial $L$ se necesita un tiempo $L/c_s$ para establecerse en un estado coherente de oscilaci\'on en el que la
densidad del plasma aumenta y disminuye de la misma manera a lo largo
de la regi\'on. Esto conduce a un l\'{\i}mite superior para la
extensi\'on espacial de cualquier oscilaci\'on ac\'ustica en el
universo primitivo. La raz\'on para la existencia de este borde es que
solo hubo un tiempo limitado, los 380~mil a\~nos antes mencionados,
para que estas oscilaciones se excitaran en el plasma
c\'osmico. Despu\'es de ese per\'{\i}odo de tiempo, las
part\'{\i}culas del plasma se combinaron para formar \'atomos
estables. Dado que el fuerte acoplamiento electromagn\'etico entre
fotones y materia depend\'{\i}a de la presencia de cargas el\'ectricas
libres (los fotones son constantemente absorbidos y reemitidos por
part\'{\i}culas cargadas), la formaci\'on de \'atomos nos lleva a que el fuerte acoplamiento de fotones y materia llegara a su fin. Hubo una ca\'{\i}da abrupta de la presi\'on y cesaron las oscilaciones.

Considerando el tiempo c\'osmico en que se formaron los \'atomos es
sencillo estimar que las oscilaciones
coherentes m\'as grandes posibles ten\'{\i}an una extensi\'on espacial
de unos 228 mil a\~nos luz (o 70 mil parsec). Simplemente no hab\'{\i}a
tiempo para m\'as: con la velocidad del sonido al 60\% de la velocidad
de la luz y un tiempo de aproximadamente 380~mil a\~nos, las regiones
m\'as grandes en las que podr\'{\i}an desarrollarse oscilaciones coherentes
ten\'{\i}an una extensi\'on espacial de $0,6 \times 380~{\rm kyr} \sim
230$ mil a\~nos luz. A este l\'{\i}mite superior se lo conoce como el ``horizonte ac\'ustico.''

\begin{figure}[tb] 
    \postscript{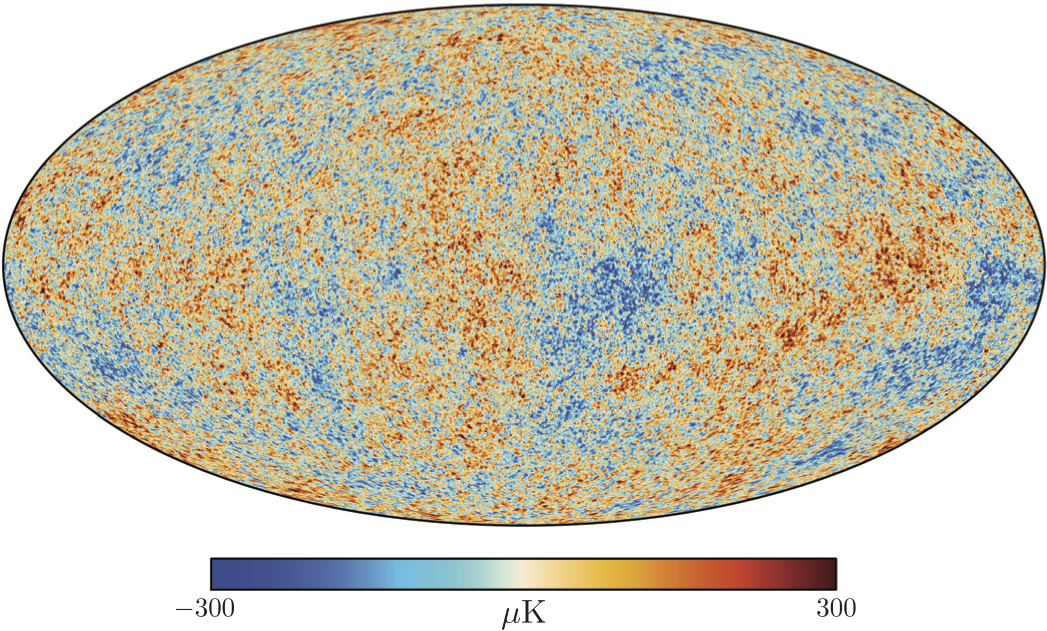}{0.99} 
    \caption{Las anisotrop\'{\i}as del FCM seg\'un lo observado por la
      misi\'on Planck de la ESA. El FCM es una instant\'anea de la luz
      m\'as antigua de nuestro cosmos, impresa en el cielo cuando el
      universo ten\'{\i}a solo 380 mil a\~nos. La figura muestra
      peque\~nas fluctuaciones de temperatura que se corresponden con
      regiones de densidades ligeramente diferentes. Cr\'edito: ESA, Planck Collaboration.
\label{fig:1}}
   \end{figure}

La nucleos\'{\i}ntesis primordial y el fondo c\'osmico de microondas
(FCM) constituyen los puntos de referencia m\'as antiguos que pueden
ser verificados observacionalmente. Nos proporcionan un {\it
  experimento} para confrontar las predicciones de $\Lambda$MOF~\cite{Jungman:1995bz}. En
efecto, usando el modelo de $\Lambda$MOF como se hubiera visto
poco antes de  la \'ultima dispersi\'on de fotones es posible predecir el
tama\~no de las fluctuaciones f\'{\i}sicas en el plasma. Las predicciones del tama\~no de estas fluctuaciones f\'{\i}sicas pueden ser comparadas con observaciones del espectro de potencia
angular de las fluctuaciones de temperatura en el FCM, que se muestran
en la Fig.~\ref{fig:1}. Al hacer que la comparaci\'on
funcione podemos calibrar los par\'ametros libres del modelo
$\Lambda$MOF (que incluyen el tama\~no {\it preciso} del horizonte ac\'ustico, la densidad bari\'onica y
la densidad de MOF) y tambi\'en podemos verificar algunas de las
suposiciones. Una vez que todos los par\'ametros fueron calibrados,
permitimos que el modelo evolucione
como la f\'{i}sica nos dice que deber\'{\i}a.
Todo este proceso
nos permite pronosticar la historia de expansi\'on del universo de acuerdo
a $\Lambda$MOF y culmina con la 
predicci\'on sobre una propiedad fundamental del universo: que tan r\'apido deber\'{\i}a
expandirse el universo al d\'{\i}a de hoy, lo que llamamos la constante de Hubble: $H_0 =
67,4 \pm 0,5~{\rm km} \, {\rm s}^{-1} \, {\rm Mpc}^{-1}$~\cite{Aghanim:2018eyx}. Es
importante tener en cuenta que
kil\'ometros y megaparsecs son unidades de longitud, por lo que el
producto de uno con el inverso del otro es
un n\'umero adimensional y, por lo tanto, la constante de Hubble tiene
dimensiones de tasa de expansi\'on por unidad de tiempo.\footnote{Un
  megaparsec es una unidad de distancia que equivale a 3,26 millones
  de a\~nos luz, una magnitud conveniente cuando consideramos la
  estructura a gran escala del universo.}  La predicci\'on de la
constante de Hubble que nos provee el modelo de $\Lambda$MOF ajustado
a las mediciones de Planck  es incre\'{\i}blemente precisa. Por lo
tanto, una prueba de principio a fin muy eficaz de toda esta historia
del modelo cosmol\'ogico $\Lambda$MOF y en particular de las
componentes del sector oscuro en los supuestos que hicimos es tratar de medir la tasa de expansi\'on del universo con una presici\'on comparable.

La expansi\'on del universo hace que todas las galaxias se alejen de
un observador dado (por ejemplo, en la Tierra), y cuanto m\'as lejos
est\'an, más r\'apido se alejan. Dicho de otro modo, las galaxias  se
alejan unas de otras con una velocidad proporcional a la distancia
entre ellas~\cite{Hubble}. La llamada ley de Hubble describe la relaci\'on entre
la distancia $d$ a un objeto dado y su velocidad de recesi\'on $v$:
\begin{equation}
  H_0 = \frac{\mathrm{velocidad \ de \  recesion}^{\!
      \!\!\!\!\!\!\! \prime}~~~}{{\rm distancia \ al \ objeto}} = \frac{v}{d} \, .
\end{equation}
Ahora, un punto que vale la pena se\~nalar en esta coyuntura es que
las galaxias no siguen exactamente la ley de Hubble. Adem\'as de la
expansi\'on del universo, los movimientos de las galaxias se ven
afectados por la gravedad de estructuras cercanas, como por ejemplo la
atracci\'on entre la V\'{\i}a L\'actea y la galaxia de Andr\'omeda. Por lo
tanto, cada galaxia tiene una velocidad peculiar, donde se usa esta
identificaci\'on en el sentido de ``individuo'' o ``espec\'{\i}fico de
s\'{\i}  mismo.'' Por lo tanto, la velocidad de recesi\'on de una
galaxia est\'a dada por la relaci\'on 
\begin{equation}
  v = H_0 \times d + v_{\rm pec} \,,
\end{equation}
donde $v_{\rm pec}$ es la velocidad peculiar de la galaxia a lo largo
de la l\'{\i}nea de la visi\'on. Si las velocidades peculiares pudieran
tener cualquier valor, entonces esto har\'{\i}a la ley de Hubble
irrelevante. Sin embargo, las velocidades peculiares son t\'{\i}picamente de unos 300~kil\'ometros por segundo, y rara vez
superan los mil kil\'ometros por segundo. Por lo tanto, la ley de
Hubble se vuelve precisa para las galaxias muy muy lejanas, cuando el
producto entre $H_0$ y $d$ es mucho m\'as grande que mil kil\'ometros
por segundo.

\begin{figure}[tb] 
    \postscript{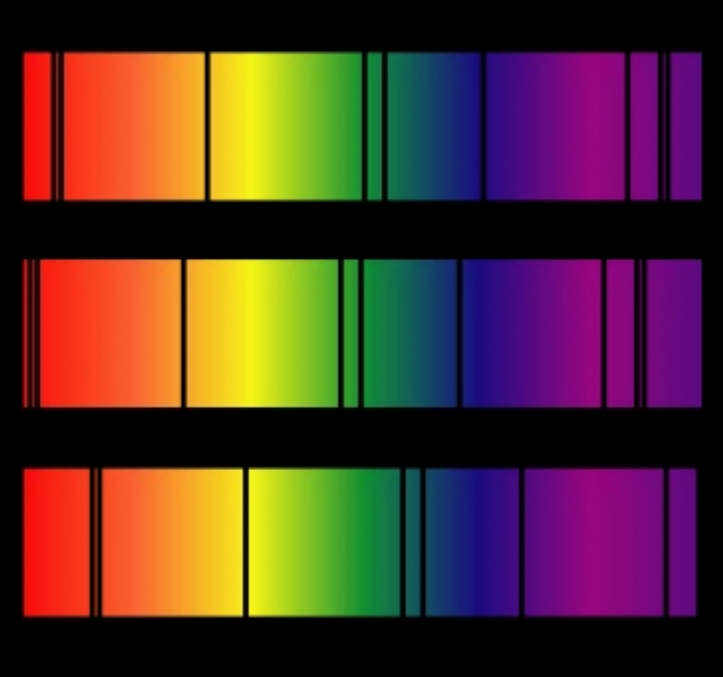}{0.99} 
    \caption{El panel superior muestra las posiciones normales de la
      l\'{\i}neas espectrales (oscuras) de una fuente en reposo. El panel central muestra el
      espectro cuando la fuente se aleja del observador. Las
      l\'{\i}neas espectrales se ven desplazadas hacia la izquierda
      (corrimiento al rojo). El panel inferior muestra el
espectro cuando la fuente se mueve hacia el observador. Las l\'{\i}neas
espectrales se ven desplazadas hacia la derecha (corrimiento al
azul). Cr\'edito: NASA.}
    \label{fig:2}
\end{figure}

La velocidad de recesi\'on de una galaxia puede medirse usando  el
llamado ``corrimiento al rojo'' de la luz que emite. En un universo en
expansi\'on las ondas de luz se {\it estiran} (aumentando as\'{\i} su
longitud de onda) desplaz\'andose cada vez m\'as hacia el rojo en el
espectro electromagn\'etico debido al efecto Doppler. Dicho efecto
puede ser medido al tomar la luz emitida por una galaxia y descomponerla en un
arcoiris para medir la cantidad de luz con que contribuye cada color
(a cada color le corresponde una longitud de onda, y una frecuencia,
diferente). Debido a que conocemos los elementos qu\'{\i}micos que
est\'an presentes en las galaxias, tales como hidr\'ogeno y ox\'{\i}geno, sabemos que \'estas emiten luz en determinadas
longitudes de onda.  As\'{\i} es que en el espectro observado de las
estrellas, las longitudes de onda ca-racter\'{\i}sticas aparecen
desplazadas al rojo con respecto al espectro original (ver Fig.~\ref{fig:2}), pudiendo as\'{\i} determinar la velocidad de recesi\'on de las estrellas y/o de las de las galaxias en las que residen. La raz\'on del cambio en la longitud de onda debido al corrimiento Doppler se llama redshift o corrimiento al rojo, y las galaxias con un alto corrimiento al rojo tienen una alta velocidad de recesi\'on.

La determinaci\'on de la distancia a las galaxias lejanas es m\'as
compleja. Con el transcurso de los a\~nos, se han encontrado
diferentes estimadores de distancias. Uno de ellos es una clase de
estrellas conocidas como variables Cefeidas. Las Cefeidas son
estrellas gigantes pulsantes que se pueden ver en galaxias
distantes. Son cien veces m\'as luminosas que el sol y la frecuencia
con la que pulsan se correlaciona muy fuertemente con su
luminosidad. En efecto, existe una fuerte correlaci\'on entre el brillo
intr\'{\i}nseco y el per\'{\i}odo de pulsaci\'on de las estrellas
variables Cefeidas: estrellas intr\'{\i}nsecamente m\'as brillantes
poseen per\'{\i}odos de variaci\'on m\'as largos. De este modo,
observando el per\'{\i}odo de cualquier Cefeida, se puede deducir su
brillo intr\'{\i}nseco y as\'{\i}, observando su brillo aparente,
calcular su distancia. De esta forma pueden usarse las estrellas
variables Cefeidas como una de las ``candelas est\'andar'' del universo,
tanto actuando como indicadores de distancia directamente, como
as\'{\i} 
tambi\'en pudiendo ser usadas para calibrar (o seleccionar el punto
cero de) otros indicadores de distancias. El nombre Cefeida proviene de la estrella
$\delta$-Cephei en la \mbox{constelaci\'on} de Cefeo, la cual fue el primer ejemplo conocido de este particular tipo de estrellas variables y es un objeto f\'acilmente visible a simple vista.

\begin{figure}[tb] 
    \postscript{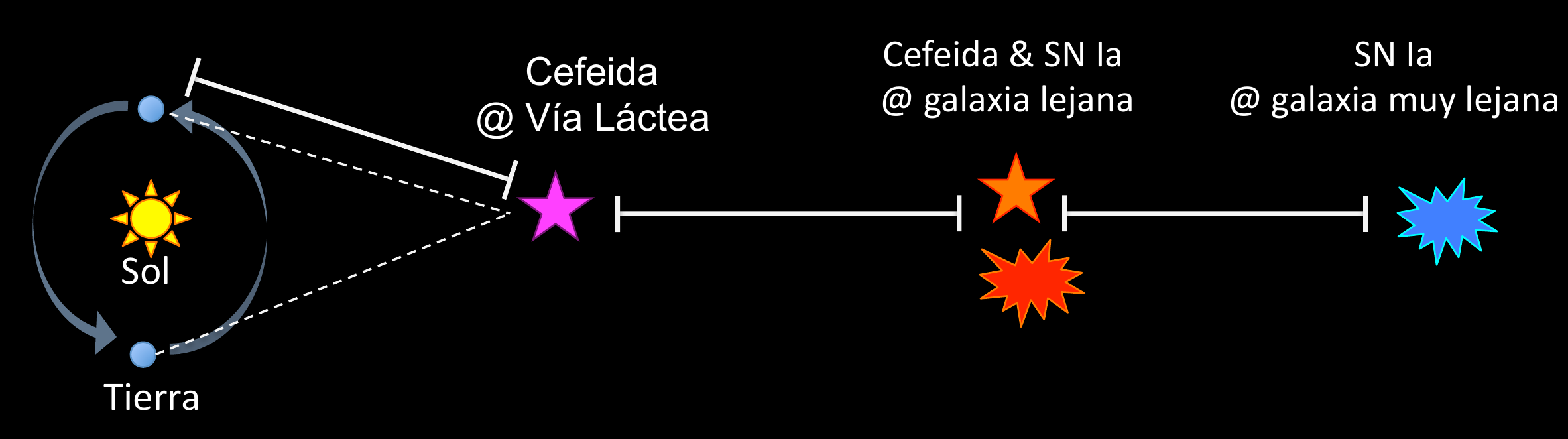}{0.99} 
    \caption{Representaci\'on esquem\'atica de la escalera de
      distancia. \label{fig:3}}
    \end{figure}

Para medir la distancia a galaxias muy lejanas el m\'etodo m\'as
    preciso es conocido como escalera de distancia. La idea es
    construir una escalera de distancia simple con escalones
    s\'olidos. El primer escal\'on es puramente geom\'etrico, basado
    en el m\'etodo de paralaje para medir la distancia a las variable
    Cefeidas en nuestra V\'{\i}a L\'actea (o en galaxias vecinas). Este
    escal\'on, que generalmente se corresponde con escalas de
    kiloparsec o megaparsec, puede interpretarse como el trazado de
    una cinta m\'etrica a estrellas cercanas.  Ahora bien, en algunas
    galaxias podemos ver tanto Cefeidas como supernovas de tipo Ia
    (SNs Ia). Las SNs Ia se producen cuando una enana blanca, el
    ``cad\'aver'' de una estrella similar al Sol, absorbe material de
    una estrella compa\~nera y alcanza una masa cr\'{\i}tica (llamada
    l\'{\i}mite de Chandrasekhar que es equivalente a 1,39 masas
    solares), lo que desencadena una explosi\'on cuya luminosidad
    ser\'a, dado su origen, similar en casi todos los casos. Esta
    uniformidad y el hecho de que el brillo alcanza unos mil millones
    de luminosidades solares convirti\'o a las SNs Ia
    en los objetos id\'oneos para medir distancias a galaxias muy
    lejanas.  Para fabricar nuestro segundo escal\'on entonces
    observamos las galaxias que recientemente albergaron una SN Ia y buscamos variables Cefeidas en esas galaxias. La
    \'unica suposici\'on para fabricar este escal\'on es que las
    Cefeidas y las SNs Ia se encuentran en el mismo lugar, y por
    lo tanto ambas esten a la misma distancia. Esta calibraci\'on
    generalmente se hace en galaxias que est\'an entre 10 y 40
    megaparsec de distancia. En el tercer y \'ultimo escal\'on, se
    calibran las SNs Ia que se encuentran en galaxias
    muy distantes (en el llamado flujo del Hubble), por lo que su
    desplazamiento hacia el rojo (que es aproximadamente 0,1) nos
    permite determinar $H_0$. El uso de SNs Ia como ``candelas
    est\'andar'' nos permite medir distancias de hasta gigaparsecs. El
    ciclo de todo este proceso que permite la medici\'on de distancias
    astron\'omicas se muestra gr\'aficamente en la Fig.~\ref{fig:3}.

En resumen, la escalera de distancia nos permite derivar un resultado emp\'{\i}rico, ya que no hay ning\'un tipo de {\it f\'{\i}sica}
involucrada; es decir no hay suposiciones del modelo astrof\'{\i}sico
ni tampoco del modelo cosmol\'ogico $\Lambda$MOF.  Siempre que las
mediciones de distancia se obtengan de una manera consistente, al propagar
tambi\'en de manera precisa los errores estad\'{\i}sticos y
sistem\'aticos (y en particular la covarianza entre los errores),
podemos obtener un valor preciso de $H_0$ para comparar con la
predicci\'on de $\Lambda$MOF. Usando el m\'etodo de la escalera de
distancias el grupo SH0ES ha observado un valor $H_0 =  74,03\pm
1,42~{\rm km} \, {\rm s}^{-1} \, {\rm Mpc}^{-1}$~\cite{Riess:2019cxk}, que se
encuentra a m\'as de $5,3\sigma$ (o desviaciones  est\'andar) de la predicci\'on de $\Lambda$MOF~\cite{Verde:2019ivm}.

\begin{figure}[tb] 
    \postscript{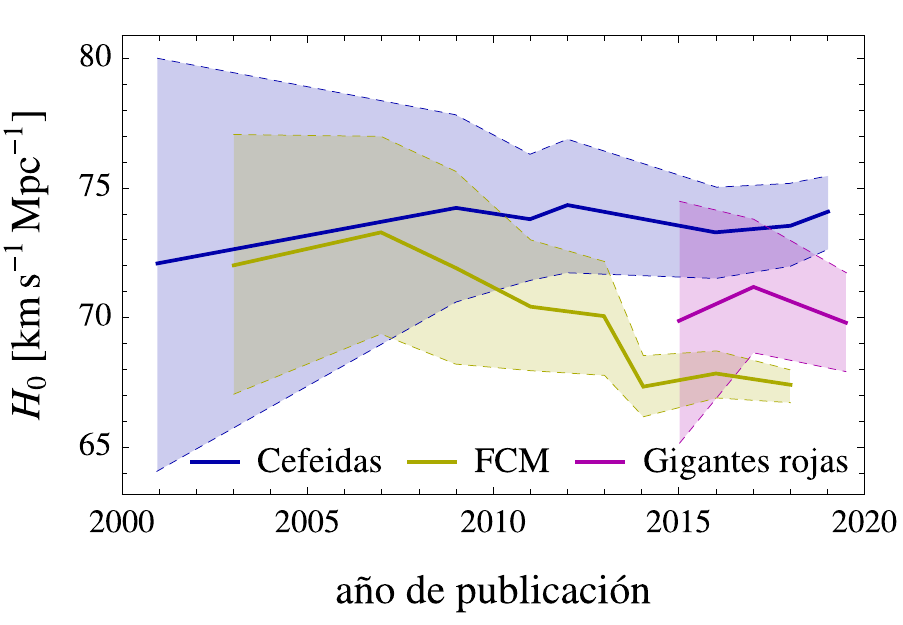}{0.99} 
\caption{Mediciones de la constante de Hubble en funci\'on de la fecha de publicaci\'on. Las l\'{\i}neas continuas indican la evoluci\'on de la media. Las regiones sombreadas abarcan valores dentro de una
      desviaci\'on est\'andar de la media. El color azul representa valores
      de $ H_0 $ determinados en el universo cercano con una calibraci\'on
      basado en la escala de distancia a Cefeidas aplicada a SNe Ia. La primera medida es de
      Hubble Key Project~ \cite{Freedman:2000cf}, las siguientes dos mediciones
      son del grupo SH0ES~\cite{Riess:2009pu,Riess:2011yx}, la tercera medici\'on es del programa Carnegie Hubble que utiliz\'o datos del infrarrojo para recalibrar los datos
del Hubble Key Project~\cite{Freedman:2012ny}, y las \'ultimas tres
mediciones son tambi\'en del grupo SH0ES~\cite{Riess:2016jrr,Riess:2018byc,Riess:2019cxk}. El color marr\'on
indica valores derivados de $ H_0 $ basados ​​en el modelo $\Lambda $MOF y medidas del FCM. Las primeras cinco medidas son
de la sonda espacial WMAP~\cite{Spergel:2003cb, Spergel:2006hy,Komatsu:2008hk,Komatsu:2010fb,Hinshaw:2012aka},
los dos siguientes son de la misi\'on
Planck~\cite{Ade:2013zuv,Ade:2015xua}, luego hay una estimaci\'on
usando el mapeo de energ\'{\i}a oscura~\cite{Abbott:2017smn}, y el \'ultimo
el punto tambi\'en pertenece a la misi\'on Planck~\cite{Aghanim:2018eyx}. El
color rojo indica medidas locales de $ H_0 $  basadas en una calibraci\'on de la rama de gigantes rojas aplicada a
SNs Ia~\cite{Jang:2015,Jang:2017dxn,Freedman:2019jwv}. 
\label{fig:4}}
\end{figure}

El ``punto de la rama de gigantes rojas'' (PRGR) es un conjunto de
estrellas que se encuentran en un punto crucial en su evoluci\'on. Las
estrellas que se encuentran en la llamada ``rama de las
gigantes rojas'' son estrellas que casi han agotado el hidr\'ogeno en
sus n\'ucleos. La siguiente etapa de su vida se desencadena cuando
comienzan a fusionar helio en sus n\'ucleos. Las estrellas en el PRGR
son las que acaban de comenzar esta etapa de quema de helio, y se
pueden distinguir por su enrojecimiento y brillo
caracter\'{\i}sticos. Estas particularidades del conjunto PRGR lo
hacen muy adecuado para medir distancias, ya que sabemos cu\'an
brillante sus estrellas deben aparecer a cierta distancia. La medici\'on m\'as reciente de la constante de Hubble, $H_0 = 69,8 \pm
1,9~{\rm km \, s^{-1} \, Mpc^{-1}}$, esta basada en una \mbox{calibraci\'on} de la rama de gigantes
rojas (o m\'as precisamente del conjunto PRGR) aplicada a SNs Ia~\cite{Freedman:2019jwv}.  El
valor obtenido es compatible (a $1,2\sigma$) con el valor estimado por
el modelo cosmol\'ogico $\Lambda$MOF. A pesar de usar SNs Ia, este m\'etodo
es independiente del m\'etodo de la escalera de distancias que usa
Cefeidas y SNs Ia. El valor de $H_0$ que resulta de la calibraci\'on a
Cefeidas se encuentra tambi\'en a menos de $2\sigma$ del
nuevo valor. En la Fig.~\ref{fig:4} se muestra la evoluci\'on de las
mediciones de la constante de Hubble en funci\'on del a\~no de
publicaci\'on. Es importante notar que tambi\'en se han llevado a cabo
varias mediciones independientes de $H_0$, aunque utilizando m\'etodos
un tanto 
dependientes del modelo astrof\'{i}sico~\cite{Wong:2019kwg}. La
tensi\'on entre las mediciones y las predicciones de $H_0$ se extiende
al  estudio del conjunto universal de datos estad\'{\i}sticamente independientes
que muestra una discrepancia de
$4,4\sigma$~\cite{Verde:2019ivm}.

La discrepancia entre que tan r\'apido parece expandirse el universo y
que tan r\'apido esperamos que se expanda ha puesto al modelo
est\'andar de la cosmolog\'{\i}a moderna en jaque. Podr\'{\i}a ser que
la tensi\'on entre el valor observado y predicho de $H_0$ no sea m\'as
que un error de medici\'on. Pero si la discrepancia es real, esta pareciera indicar que el universo primitivo se comport\'o
de manera diferente a la predicha por $\Lambda$MOF. Ahora bien,
\textquestiondown c\'omo
podemos solucionar el problema? Est\'a claro que una dosis extra de
radiaci\'on en el universo primitivo podr\'{\i}a conciliar los valores
en conflicto de la constante de Hubble, ya que la presi\'on externa de
esta radiaci\'on habr\'{\i}a acelerado la expansi\'on del universo
antes de que se formara la luz observada del FCM. 

Al principio de la discusi\'on sugerimos que 
la manera m\'as simple de explicar la energ\'{\i}a oscura es la
constante cosmol\'ogica $\Lambda$, la energ\'{\i}a del espacio mismo,
con una densidad constante en todas partes. Pero, \textquestiondown qu\'e ocurr\'{\i}a
si la cantidad de energ\'{\i}a oscura en el universo no es constante?
Un breve per\'{\i}odo dominado por la energ\'{\i}a oscura en el
universo primitivo (llamada energ\'{\i}a oscura temprana) en principio podr\'{\i}a
reconciliar las mediciones de $H_0$~\cite{Poulin:2018cxd}. De manera alternativa, uno podr\'{\i}a postular la existencia de tres
sabores de neutrinos dextr\'ogiros $\nu_R$ junto con sus antineutrinos
lev\'ogiros $\bar \nu_L$ para restaurar la simetr\'{\i}a quiral del modelo est\'andar de la f\'{\i}sica de part\'{i}culas~\cite{Anchordoqui:2011nh}. Sin embargo, tambi\'en debemos tener en
cuenta que el agregado de part\'{\i}culas relativistas puede distorcionar
la abundacia de los elementos que fueron sintetizados durante la
nucleos\'{\i}ntesis primordial~\cite{Steigman:1977kc}. En particular, la abundancia de n\'ucleos
de helio-4 depende de la abundancia relativa de neutrones $n$ y
protones $p^+$ al momento de la s\'{\i}ntesis del helio.  Cuando el
universo ten\'{\i}a menos de un segundo de edad, los protones y los
neutrones pod\'{\i}an absorber y emitir neutrinos libremente para
transformarse entre s\'{\i}. La continua conversi\'on de protones en
neutrones y viceversa estaba gobernada por los siguientes procesos de
la llamada interacci\'on d\'ebil,
\begin{equation}
  p^+ + \bar \nu \leftrightharpoons n + e^+ \quad \quad y \quad \quad
  p^+ + e^- \leftrightharpoons n + \nu \,,
\end{equation}
donde $e^-$ denota un electr\'on y $e^+$ su antipart\'{\i}cula el
positr\'on, y donde la doble flecha indica que la reacci\'on puede
suceder en ambos sentidos. Estas transmutaciones eran igual de r\'apidas en cualquier direcci\'on, por
lo que en el plasma primordial hab\'{\i}a la misma cantidad de
protones que de neutrones. Pero a medida que la temperatura del
universo disminuy\'o,
la energ\'{\i}a se volvi\'o escasa para mantener las interacciones
d\'ebiles en equilibrio. Debido a que se necesita m\'as energ\'{\i}a
para producir un neutr\'on que un prot\'on, que es un poco menos
masivo, la cantidad de neutrones comenz\'o a disminuir en relaci\'on
con la cantidad de protones. Cuando el universo alcanz\'o la edad de un
segundo, y su temperatura estaba por debajo de los 10 mil millones de
grados kelvin, la expansi\'on se hizo tan r\'apida que las
transmutaciones entre las poblaciones (ahora desiguales) de protones y
neutrones no pudieron mantener el ritmo.  Esto dej\'o una proporci\'on
particular de neutrones a protones en el universo primitivo. Tres
minutos despu\'es, las reacciones nucleares los transformaron en una
abundancia c\'osmica definida de helio. De este modo, la abundancia que hoy observamos de helio-4 depende de manera crucial de la
proporci\'on entre protones y neutrones antes de que comiencen las
reacciones nucleares. Esto, a su vez, depende de que tan r\'apido se
estaba expandiendo el universo en ese momento. Y aqu\'{\i} es donde la
cantidad de neutrinos entra en la historia. Cada especie de neutrino
relativista se suma a la densidad total del universo y aumenta la
velocidad de expansi\'on. Cuanto m\'as r\'apida sea la expansi\'on,
m\'as r\'apido se terminan las transmutaciones y mayor ser\'a el
n\'umero de neutrones sobrevivientes. M\'as neutrones
significar\'{\i}an una mayor abundancia de helio-4. Combinando datos
del FCM y mediciones de la abundancia de helio-4 es posible obtener
l\'imites a la cantidad de radiaci\'on que uno puede agregar en el
universo primitivo~\cite{Aghanim:2018eyx}. Dichos l\'{\i}mites ponen severas restricciones a
los modelos alternativos a $\Lambda$MOF~\cite{Anchordoqui:2019yzc}.

Los datos colectados
hasta el momento sugieren que la soluci\'on m\'as
convincente pareciera  ser la existencia de tres 
$\nu_R + \bar \nu_L$.  Sin embargo, si las mediciones futuras del par\'ametro Hubble tienen
barras de error m\'as peque\~nas pero con el mismo valor central, los modelos
de energ\'{i}a oscura temprana podr\'{\i}an verse favorecidos ya que
permiten valores mayores para $H_0$~\cite{Agrawal:2019lmo}. La
soluci\'on al problema de $H_0$ tambi\'en podr\'{\i}a manifestarse en
un cambio en la tasa de expansi\'on actual,
como por ejemplo introduciendo una nueva interacci\'on entre la energ\'{\i}a y
la materia oscura~\cite{Agrawal:2019dlm,DiValentino:2019exe}. Sin
embargo, a\'un teniendo en consideraci\'on todas las libertades adicionales mencionadas, la mayor\'{\i}a de los modelos cosmol\'ogicos
alternativos a $\Lambda$MOF solo reducen la tensi\'on en la constante de Hubble en
lugar de eliminarla~\cite{Benevento:2020fev}. Predicen una tasa de expansi\'on c\'osmica m\'as
r\'apida que $\Lambda$MOF, pero a\'un no es lo suficientemente
r\'apida como para igualar las observaciones de supernovas y otros
objetos astron\'omicos. En principio, con modelos combinados uno
podr\'{\i}a encontrar una 
soluci\'on global~\cite{Anchordoqui:2019amx,Anchordoqui:2020sqo}.

En los pr\'oximos a\~nos, el telescopio Euclides  mapear\'a
meticulosamente c\'omo la gravedad y la energ\'{\i}a oscura han moldeado
la evoluci\'on c\'osmica~\cite{Laureijs:2011gra}. Los nuevos datos experimentales tal
vez nos permitan descartar definitivamente al modelo $\Lambda$MOF. Las
observaciones de Euclides tambi\'en nos permitir\'an hacer un testeo
exhaustivo y mucho 
m\'as restrictivo de todos los modelos alternativos propuestos.\\

Carlos Garc\'{\i}a Canal someti\'o el manuscrito final a una
cr\'{\i}tica exhaustiva, por lo cual le doy las gracias. L.A.A.  es subvencionado por U.S. NSF (Grant PHY-1620661) y NASA (Grant
  80NSSC18K0464).

\end{document}